# Quantifying Hidden Architectural Patterns in Metaplastic Tumors by Calculating the Quadrant-Slope Index (QSI)

David H. Nguyen, Ph.D.
Affiliate Scientist
Dept. of Cellular & Tissue Imaging
Division of Molecular Biophysics and Integrated Bioimaging
Lawrence Berkeley National Laboratory
DHNguyen@lbl.gov

April 2017

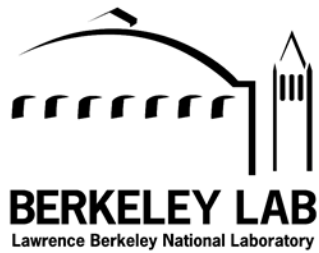

## Abstract
The Quadrant-Slope Index (QSI) method was created in order to detect subtle patterns of organization in tumor images that have metaplastic elements, such as streams of spindle cells [1]. However, metaplastic tumors also have nuclei that may be aligned like a stream but are not obvious to the pathologist because the shape of the cytoplasm is unclear. The previous method that I developed, the Nearest-Neighbor Angular Profile (N-NAP) method [2], is good for detecting subtle patterns of order based on the assumption that breast tumor cells are attempting to arrange themselves side-by-side (like bricks), as in the luminal compartment of a normal mammary gland [3]. However, this assumption is not optimal for detecting cellular arrangements that are head-to-tail, such as in streams of spindle cells. Metaplastic carcinomas of the breast (i.e. basal-like breast cancers, triple-negative breast cancers) are believed to be derived from the stem or progenitor cells that reside in the basal/myoepithelial compartment of the normal mammary gland [Reviewed in 3]. Epithelial cells in the basal/myoepithelial compartment arrange themselves in an head-to-tail fashion, forming a net that surrounds the luminal compartment [3,4]. If cancer cells in a metaplastic tumor are trying to be normal, the optimal way to detect subtle regions of them attempting to be ordered normally should highlight the head-to-tail alignment of cells.

## Graphical Abstract

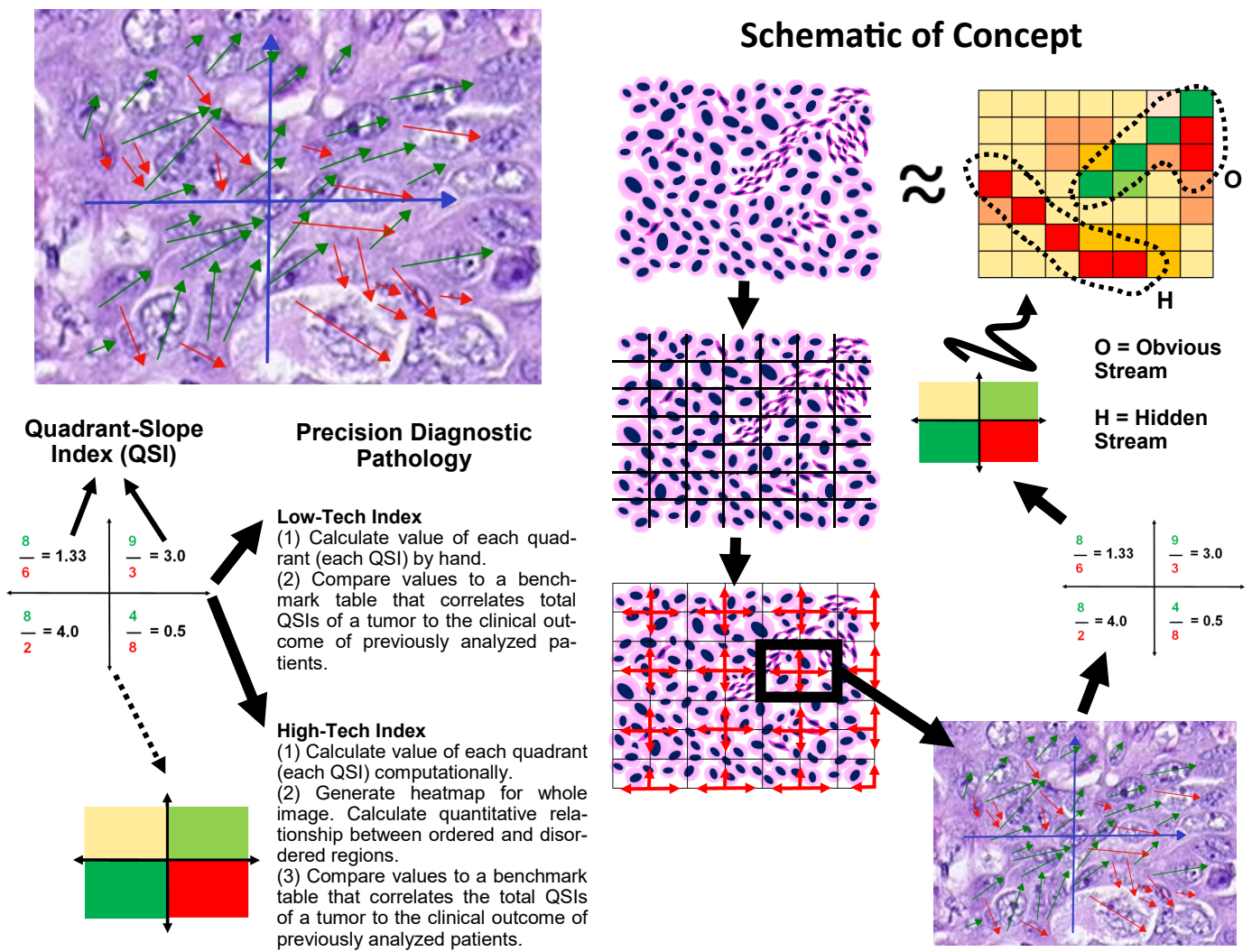

# Figure 1

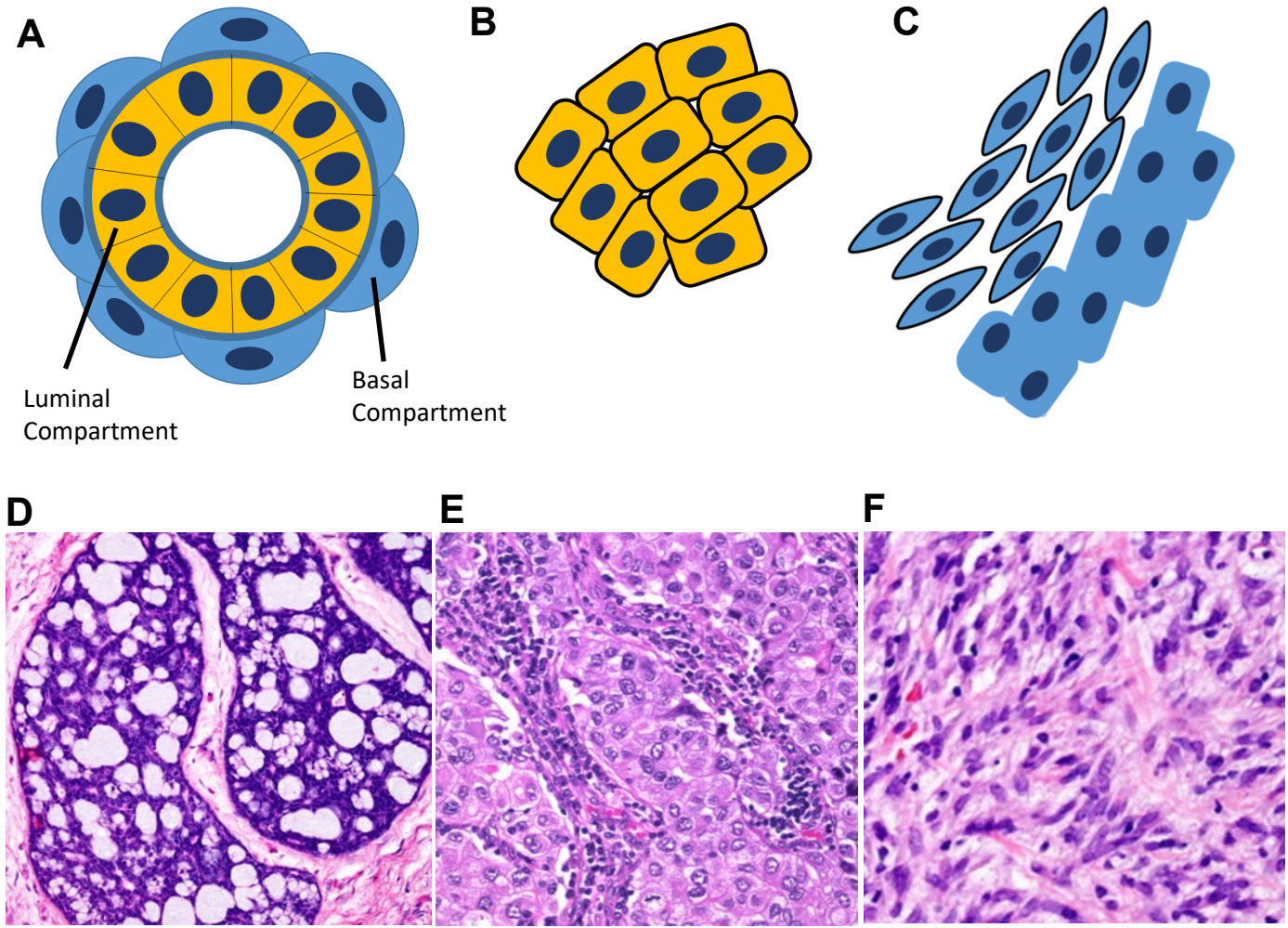

**Figure 1. Rationale Behind Quadrant-Slope Index (QSI) for Detecting Head-to-Tail Orientation.**
(A) Schematic of a normal mammary duct containing a luminal compartment consisting of epithelial cells that pack side-by-side like bricks to form a wall. The basal or myoepithelial compartment consists of cells that are arranged head-to-tail, forming a net that surrounds the luminal compartment and has contractile functions [3].
(B) Schematic of tumor cells that have a luminal nature. They attempt to pack side-by-side, like bricks. The nearest-neighbor angular profile (N-NAP) method is great for detecting side-by-side nuclear arrangements.
(C) Schematic of tumors cells that have a myoepithelial nature. They attempt to pack head-to-tail, similar to a layered net [1]. The N-NAP method [2] is not optimal for detecting head-to-tail arrangements inside tumors. Thus, the Quadrant-Slope Index (QSI) method was developed.
(D-F) Examples of glandular breast tumors and a spindle cell breast tumor. Tumors can exhibit seemingly random arrangement of nuclei. The magnification of each image is varied in order to highlight distinct nuclear and cellular features. Images adapted from WebPathology.com. (D, E) Morphological features of adenoid tumors that exhibit an attempt to form normal mammary tubes. (F) An example of a spindle cell carcinoma that exhibits head-to-tail arrangement of nuclei.

# Figure 2

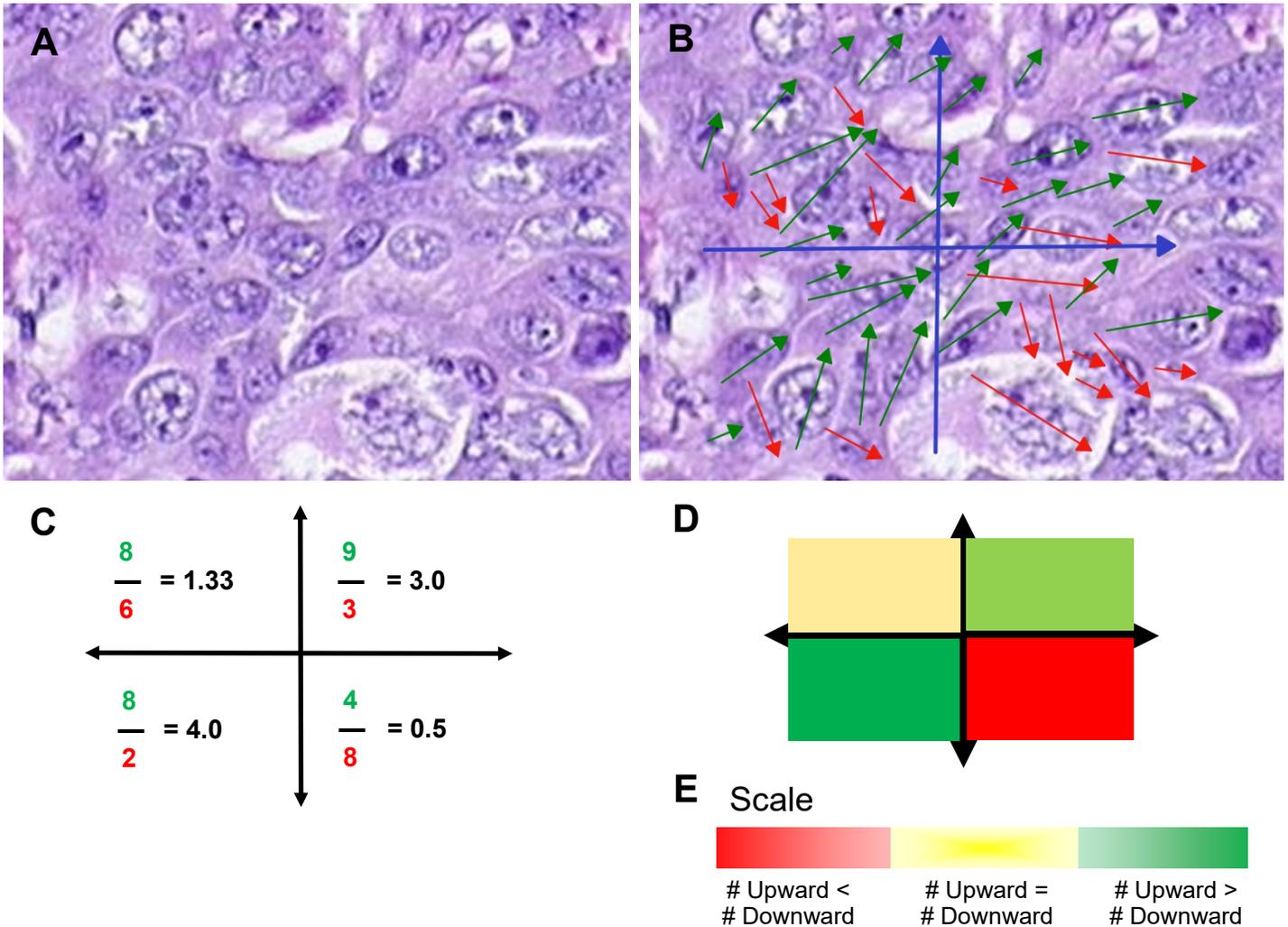

**Figure 2. How the Quadrant-Slope Index (QSI) is Calculated.**
(A) Tumors can exhibit seemingly random arrangement of nuclei.
(B) A uniformly square grid is overlaid on the image (not shown here, see Figure 3B). Each square is then divided into quadrants by placing the origin of an X,Y plane at the center of the square. A line along the longest length within each nuclei is drawn. The slope of a line, relative to the horizontal x-axis, is categorized as either "up" (green) if the line slants upwards from left to right as it moves along the positive x-axis, or "down" (red) if the line slants downwards from left to right.
(C) The QSI of each quadrant is calculated by counting the number of upward slopes and then dividing that by the number of downward slopes. Quadrants with an equal number of up or down slopes will have a QSI value of 1.00. Quadrants with more upward slopes will have a QSI value >1.00; those with more downward slopes will have a QSI value <1.00. Transforming all QSI values by y=log(x^3) will make the magnitude of the QSIs more intuitive. For example, the bottom right quadrant (quadrant IV) goes from being 0.50 to approximately -1.00. The greater the number of downward slopes (red) than upward slopes (green) in a quadrant, the closer to zero the QSI value will be on a linear scale (i.e. 1÷15 = 0.07), which is cumbersome. Transforming the linear scale value by y=log(x^3) turns 0.07 into -3.45 [y=log(0.07^3) = -3.45], which is more intuitive. Just as with the upward (green) slopes, the more of downward slopes (red) outnumber the upward slopes, the larger the QSI magnitude should be. (D,E) The QSI values in C can be represented as a heat map. This method of visualization for the whole image can reveal regions of order that are obvious or hidden (see Figure 3C). Breast tumor image is adapted from the Stanford Tissue Array database.

# Figure 3

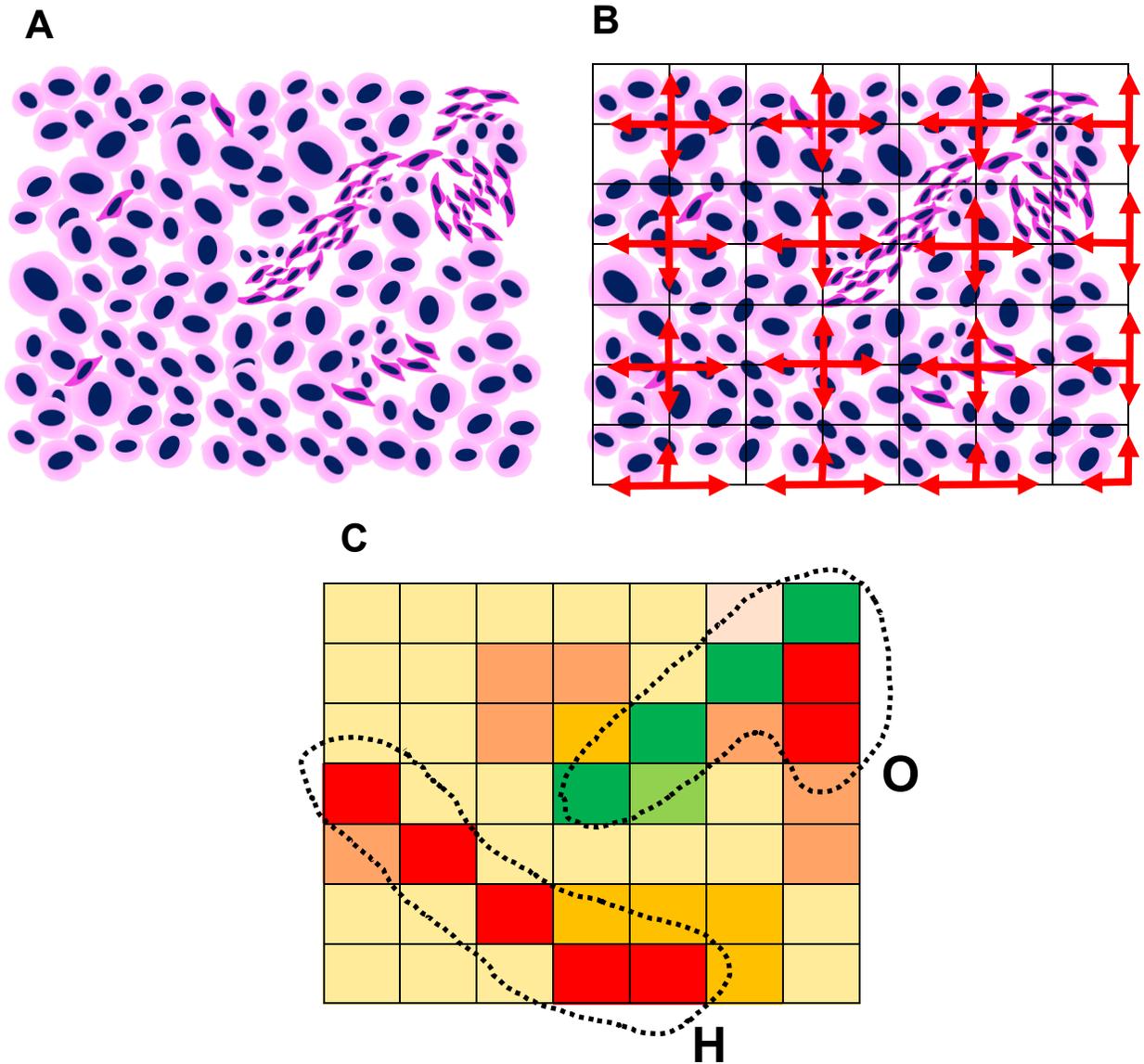

**Figure 3. Calculating the QSI for the entire tumor image can reveal regions of hidden head-to-tail order.**
(A) Schematic of a breast tumor with metaplastic elements, such as streams of spindle cells. It also contains a stream of non-spindle cells that is not obvious to the naked eye.
(B) Applying a uniform grid over the tumor in A results in units that can be further subdivided into quadrants.
(C) Visualizing the QSI of each unit in the grid can reveal obvious (O) and hidden (H) regions of head-to-tail arrangement in the tumor.

# Applications of QSI in Precision Diagnostic Pathology/Oncology

**Low-Tech Diagnostic Tool**

1.) The QSI method can be hand-calculated as a low-tech form of precision diagnostic oncology. It can supplement traditional grading methods such as nuclear morphology and staining for tissue-specific markers. However, this requires that a data table containing QSI values from previously analyzed patients to be available, such that interpolations can be made.

An important question to answer when using the QSI method as a low-tech diagnostic method is how many QSIs need to be calculated across the whole tumor image in order to provide an accurate assessment of that tumor? It is labor-intensive to calculate a QSI for every grid unit in a tumor image, which can be large.

**High-Tech Diagnostic Tool**

2.) The QSI method can be computationally calculated through image analysis pipelines. Computational approaches will make calculating the QSI for every grid unit feasible.

3.) Calculating all QSIs in a tumor allows for visualizing regions of head-to-tail alignment of nuclei. This will reveal obvious and hidden regions of order (Figure 3). This will also allow pathologists to quantify an index representing the percent of regions that are ordered, out of the whole image, for each tumor. The percent of "regions that are ordered" index may reveal subtypes within breast cancer subtypes (i.e. triple-negative, HER2) that have different rates of survival, relapse, or drug resistance.